\documentclass[pra,twocolumn,superscriptaddress,notitlepage,nofootinbib]{revtex4-2}
\usepackage[english]{babel}
\usepackage{siunitx}

\usepackage{amsmath}
\usepackage{graphicx}
\usepackage[colorlinks=true, allcolors=blue]{hyperref}
\usepackage{mathrsfs}
\usepackage{amssymb}
\usepackage{array}

\usepackage[dvipsnames]{xcolor}

\begin{document}
\title{Explaining Optomechanical Libration Spectra: A Stochastic Simulation Approach}
\author{Ankush Gogoi}
\affiliation{UCD Centre for Mechanics, School of Mechanical and Materials Engineering, University College Dublin, Dublin, Ireland}
\author{Vikram Pakrashi}
\affiliation{UCD Centre for Mechanics, School of Mechanical and Materials Engineering, University College Dublin, Dublin, Ireland}
\author{Joanna A. Zielińska}
\email{jzielinska@tec.mx}
\affiliation{School of Engineering and Sciences, Tecnológico de Monterrey, Monterrey, N.L., Mexico}
%

\begin{abstract}

In this work, we present a practical and computationally effective Ito-Taylor expansion based stochastic simulation framework for  modeling rotational optomechanics experiments. By developing a model using this framework, we could  capture the nonlinear orientation dynamics of an optically levitated, nearly cylindrically symmetric nanodumbbell. It successfully reproduces and explains shoulder-like features observed in the power spectral density of libration, which we show arising from the interplay between confined libration and thermally driven rotation around the particle’s symmetry axis.

\end{abstract}

\maketitle

\section{Introduction}

The study of the dynamics of nano and microscale objects levitated in vacuum has rapidly developed into a versatile platform for sensing weak forces and torques, testing macroscopic quantum phenomena, and searching for physics beyond the standard model~\cite{gonzalez2021levitodynamics, moore2021searching}. While control over the center-of-mass (COM) motion of particles around \qty{100}{\nano\meter} in size is steadily advancing toward the quantum regime~\cite{rossi2025delocalization}, increasing attention is being directed toward the rotational degrees of freedom of levitated anisotropic nanoparticles. Recent progress in this area includes quantum ground-state cooling of torsional oscillations (libration)~\cite{dania2025high}, coupling to internal spin degrees of freedom~\cite{delord2020spin}, record-setting torque sensitivities~\cite{ju2023near}, and interactions with the transverse orbital angular momentum of light~\cite{hu2023structured}, among other notable achievements.

In certain cases, the alignment dynamics of an optically levitated nanoparticle can closely resemble center-of-mass motion, effectively acting like harmonic oscillators. This occurs for non-cylindrically symmetric particles (typically multi-particle clusters comparable in size to the optical trap), where all three librational degrees of freedom are tightly confined, resulting in clean Lorentzian power spectral densities with minimal cross-coupling~\cite{kamba2023nanoscale, pontin2023simultaneous, gao2024PRR}. In contrast, smaller particles such as dumbbells, composed of two nanospheres, exhibit only two confined libration modes under linearly polarized optical tweezers, while the third mode remains effectively free. This results in spectral features that differ from those expected for fully harmonically confined systems.

However, even in the simplest case of nanodumbbell libration in linearly polarized optical tweezers, a variety of libration power spectral densities (PSDs) have been reported in the literature. These include single Lorentzian peaks \cite{Li23assembly}, Lorenzian peaks with shoulder-like features \cite{vanderLaan2021PRL, vanderLaan2020PRAthermalrotor, Zielinska2023PRL, zielinska2024PRL}, as well as irregular, split spectral structures \cite{seberson2019parametric, bang2020five, ju2023near}. Clearly, the shape of these spectra is highly sensitive to precise experimental conditions such as deviations from perfect cylindrical symmetry of the particle, and the exact intensity and polarization profile of the optical field gives rise to both conservative and non-conservative torques. 

The inherent nonlinearity and stochastic nature of the system make theoretical modeling of rotational degrees of freedom in levitated optomechanics particularly challenging. Accurate simulations must account for the full three-dimensional rotational dynamics of the particle, incorporating stochastic thermal torques, an often complex optical potential landscape, and nonlinear terms describing the coupling between the various degrees of freedom. 

A fundamental framework for describing the dynamics of nonlinear systems subjected to random external forces is provided by stochastic differential equations (SDEs)\cite{kloedenplaten1992,tripura20115}. Numerical integration schemes based on Ito-Taylor expansion have proven effective for solving nonlinear SDEs across a wide range of applications \cite{gogoi2022computational,tiwari2021shape,mucchielli2022mathematically}.

In this work, we introduce a computationally efficient stochastic simulation framework based on Ito-Taylor expansion that captures the nonlinear, three-dimensional orientation dynamics of levitated nearly cylindrically symmetric particles. By modeling both deterministic and thermal torques in realistic optical fields, we are able to reproduce and explain shoulder-like features observed in the libration PSD of nanodumbbell-shaped rotors. We show that these spectral signatures originate from the interplay between confined and thermally fluctuating rotation around the particle’s symmetry axis.

Our stochastic simulation approach has the potential to become essential for supporting a wide range of applications, including high-precision torque sensing~\cite{ju2023near}, the development of levitated gyroscopes~\cite{zeng2024optically}, and the study of macroscopic quantum rotational phenomena~\cite{stickler2021quantum}.



This paper is structured as follows. In Sec. II, we describe the orientational dynamics of a levitated nanoparticle. We  discuss the equations of motion and the approximations we employ. Next, in Section III, we present the stochastic simulation framework based on stochastic Ito-Taylor expansion. Throughout Section IV, we compare the simulation results with experimental data, and Section V concludes the paper.

\section{Description of the dynamics}

We first outline the physical characteristics of the nanodumbbell and introduce the formalism used to describe its rotational degrees of freedom. We then describe the orientation-dependent optical potential resulting from the particle's anisotropic geometry. Finally, we present the stochastic equations of motion governing the rotational dynamics and discuss the role of thermally fluctuating torques that drive the system.

\begin{figure}
    \centering
    \includegraphics[width=0.5\linewidth]{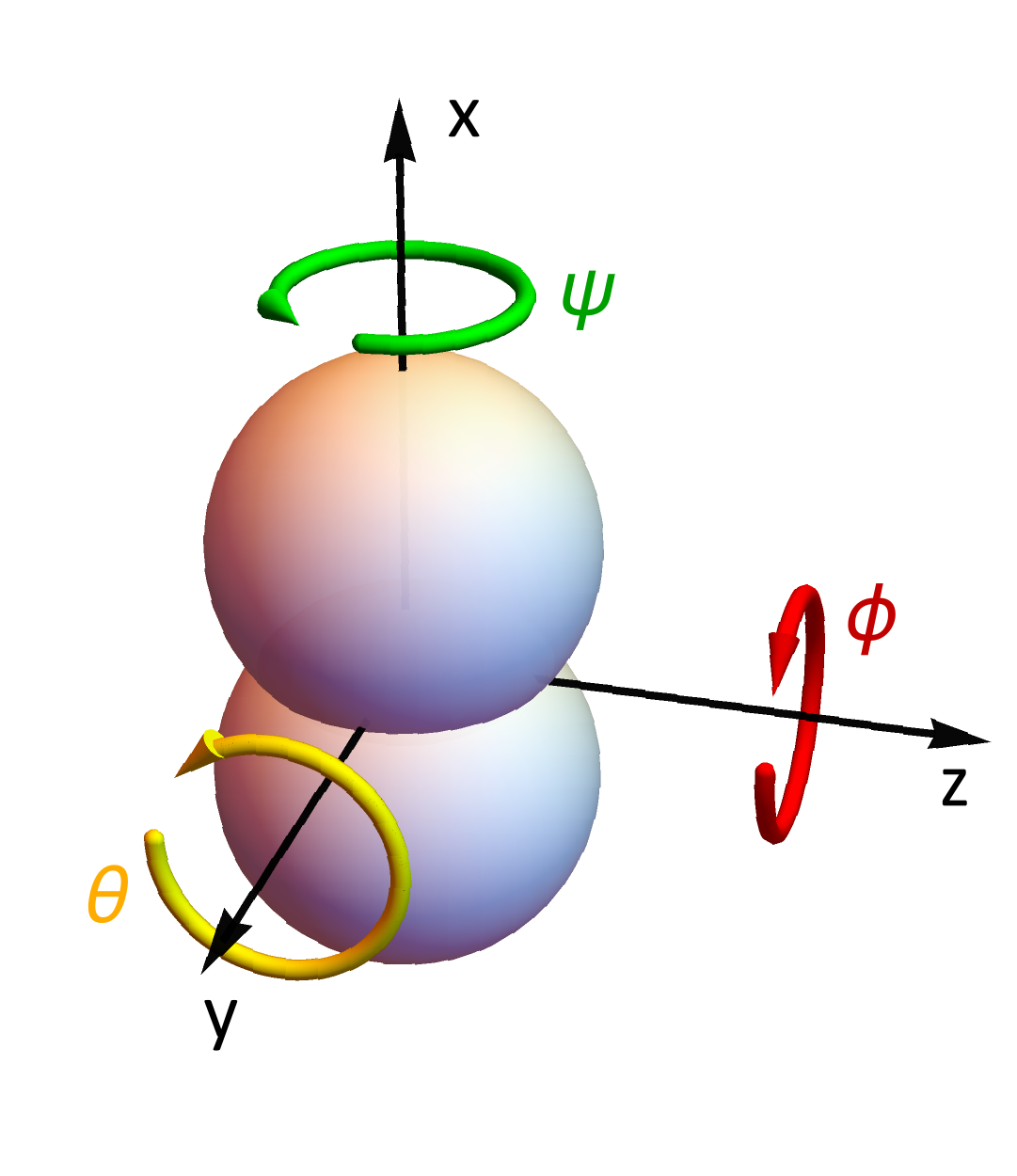}
    \caption{Rotational degrees of freedom of a nanodumbbell with its long axis aligned along the laboratory $x$-axis. In a typical levitated optomechanics setup with the optical tweezers beam propagating along the $z$-axis, the time-dependent behavior of the angle $\phi$, representing the orientation of the long axis in the $xy$-plane, is measured.}
    \label{fig:angles}
\end{figure}

This work focuses on a nearly cylindrically symmetric nanodumbbell made of two dielectric nanospheres. We denote $\boldsymbol{\alpha}=\text{diag} (\alpha_1,\alpha_2,\alpha_3)$ and $\boldsymbol{I}= \text{diag} (I_1,I_2,I_3)$ as the static electric polarizability and inertia tensors of the object in the intrinsic body frame, respectively, assuming $\alpha_1 \approx\alpha_2 \leq \alpha_3$ and $I_1 \approx I_2 \geq I_3$. We refer to the principal axis with the largest polarizability and the smallest moment of inertia as the ``long axis'' of the object. In order to describe the orientation of the body reference frame with respect to the laboratory frame ($x,y,z$), we use the intrinsic $x$-convention of Euler angles denoted as $\phi$, $\theta$ and $\psi$  (see  \cite{WolframEuler} and \S  35 in~\cite{Landau1976Mechanics}), as illustrated in Fig.~\ref{fig:angles}. The Euler angles $\phi$ and $\theta$ define the orientation of the rotor's long axis, while $\psi$ describes the rotation around the long axis, indicating the orientation of the intermediate axis, associated with the intermediate moment of inertia. In a typical levitated optomechanics experiment, with optical tweezers beam propagating along $z$, the only Euler angle accessible to measurement is $\phi$, corresponding to the orientation of the long axis in the $xy$-plane~\cite{Tebbenjohanns2022PRA}.

The nanodumbbell is trapped by an optical tweezers beam propagating along the $z$-axis and linearly polarized along the $x$-axis. We assume that the particle’s centre of mass (COM) remains fixed at the center of the optical trap. The anisotropic electric polarizability of the nanodumbbell gives rise to libration dynamics. The rotor’s long axis undergoes two-dimensional libration oscillations  around its equilibrium orientation, aligned with the polarization direction of the optical tweezers ($x$-axis). These two libration modes are described by the angles \( \phi \) and \( \theta \), and both exhibit approximately the same characteristic frequency
\begin{equation}
\Omega_0 = \sqrt{\frac{\mathcal{E}^2 (\alpha_3 - \alpha_1)}{2 I_1}},
\end{equation}
where \( \mathcal{E} \) is the electric field amplitude in the center of the trap~\cite{seberson2019parametric}. Since in thermal equilibrium we expect \( \phi \) and \( \theta \)  to perform small-amplitude oscillations, i.e.,$\sqrt{\frac{k_B T}{I_1 \Omega_0^2}} \ll 1$, we employ the small-angle approximation for \( \phi \) and \( \theta \), which is equivalent to assuming a perfectly harmonic libration potential. The third rotational degree of freedom, $\psi$, is unconstrained if the particle is perfectly cylindrically symmetric. However, a slight asymmetry (specifically, a small difference between $\alpha_1$ and $\alpha_2$) can lead to a weak confinement of the intermediate axis along $z$, resulting in libration of $\psi$. This confinement can arise from both intensity gradient torques~\cite{kamba2023nanoscale} and near-field effects due to strong focusing~\cite{novotny2012principles}. Experimental studies suggest that shape asymmetries are common in nominally spherical particles~\cite{kamba2023nanoscale, Zielinska2023PRL}. 

The system is assumed to be in thermal equilibrium and all three rotational degrees of freedom of the nanodumbbell are subject to damping and are driven by the thermally fluctuating torques arising from the surrounding environment (gas in the vacuum chamber). 

We model the dynamics of the Euler angles described above using the following equations of motion, which are derived from the Euler equations in the dumbbell’s body frame~\cite{seberson2019parametric,Landau1976Mechanics,zielinska2024PRL}:

\begin{subequations}
\label{eq:motioneq}
\begin{align}
   I_1 &\ddot{\theta}+ I_3 \dot{\phi}\dot{\psi} + I_1 \gamma_1 \dot{\theta} +I_1 \Omega_0^2 \theta= \tau_{{\rm fl},\theta} \;,\label{eq:motion21}\\
   I_1 &\ddot{\phi}- I_3 \dot{\theta}\dot{\psi} + I_1 \gamma_1 \dot{\phi} +I_1 \Omega_0^2\phi = \tau_{{\rm fl},\phi} \;,\label{eq:motion22}\\
   I_3 &\ddot{\psi}+ I_3 \gamma_3 \dot{\psi}+\frac{1}{2}I_3 \Omega_\psi^2 \sin{2\psi}=\tau_{{\rm fl},\psi},
    \label{eq:motion23}
\end{align}
\end{subequations}
where thermal fluctuating torques are denoted as $\tau_{{\rm fl},\theta},\tau_{{\rm fl},\phi}$ and $\tau_{{\rm fl},\psi}$; $\Omega_\psi$ is a parameter describing $\psi$ confinement, and $\gamma_1$ and $\gamma_3$ are damping coefficients. The two harmonically oscillating libration modes, associated with angles $\phi$ and $\theta$, are coupled, with a coupling strength proportional to $\dot\psi$. This type of coupling is characteristic of spinning top dynamics and causes the two degenerate libration modes to hybridize, splitting into distinct precession and nutation modes~\cite{seberson2019parametric}. Consequently, the dynamics of the \( \psi \) angle become encoded in the motion of the long axis and, as will be shown later, determine the libration PSD of \( \phi \) and \( \theta \).

The tight angular confinement of \( \phi \) and \( \theta \) justifies approximating their libration potentials as harmonic. However, this approximation does not hold for the weakly confined \( \psi \) angle, and therefore we retain its full sinusoidal potential in the form  
\begin{equation}
U(\psi) = \frac{1}{4} I_3 \Omega_\psi^2 \cos(2\psi).
\label{eq:psipotential}
\end{equation}
The corresponding deterministic torque term in Eq.~\eqref{eq:motion23} is derived from this potential.

The fluctuating torques which are driving the particle follow the fluctuation-dissipation theorem,
\begin{equation}
\langle \tau_{{\rm fl},i}(t_0)\tau_{{\rm fl},j}(t_0-t) \rangle = 2  I_i  \gamma_i k_{\rm B} T \delta(t) \delta_{ij},
\label{eq:torque_fluct}
\end{equation}
and are modeled as 
\begin{subequations}
\label{eq:noiseterms}
\begin{align}
   \tau_{{\rm fl},\theta} &= \sqrt{2 I_1 \gamma_1 k_B T} \bar{\eta}_\theta (t)\;,\\
   \tau_{{\rm fl},\phi} &= \sqrt{2 I_1 \gamma_1 k_B T} \bar{\eta}_\phi (t)\;,\\
   \tau_{{\rm fl},\psi} &= \sqrt{2 I_3 \gamma_3 k_B T} \bar{\eta}_\psi (t)\;,
\end{align}
\end{subequations}
where ${\eta}_\theta(t)$, $\eta_\phi(t)$ and $\eta_\psi (t)$ are independent Wiener processes such that $\bar{\eta}_\theta(t)$, $\bar{\eta}_\phi(t)$ and $\bar{\eta}_\psi (t)$ represent their corresponding generalized derivatives.

\section{Ito-Taylor expansion based stochastic integration scheme}
In this section, we first recall the definition of Ito SDEs and outline the construction of stochastic integration schemes based on Ito’s lemma and the Ito-Taylor expansion. We then present the associated multiple stochastic integrals and the generalized discretization scheme for multi-dimensional Wiener processes. Finally, we illustrate the solution of the underlying SDEs using the Ito-Taylor $1.5$ scheme.

Consider the $\mathscr{F}_t$-measurable stochastic process, $y_k\left(t\right)$ under the filtered probability space $\left(\text{S}, \mathscr{F}, (\mathscr{F}_t)_{t\ge 0}, P \right)$ such that the governing Ito SDE for the $d$-dimensional diffusion process driven by $m$-dimensional Wiener process can be expressed as \cite{gogoi2022computational,tripura20115},
\begin{align}
    \label{eq:genSDE}
   &\text{d}{y_k}\left( t \right) = {a _k}\left( {t,y\left( t \right)} \right)\text{d}t + \sum\limits_{j = 1}^m {{b _{k,j}}\left( {t,y\left( t \right)} \right)\text{d}{B^j}\left( t \right)}, \notag \\
   &\text{with}, \quad k = 1,2,\ldots d,
\end{align} 
where ${a _k}\left( {t,y\left( t \right)} \right)$, ${{b_{k,j}}\left( {t,y\left( t \right)} \right)}$ are the drift and diffusion coefficients, respectively, and ${{B^j}\left( t \right)}$ is the $\mathscr{F}_t$-measurable Wiener process with stationary independent increments following Gaussian distribution. The SDE in Eq.~\eqref{eq:genSDE} has a unique solution if the drift and diffusion coefficients satisfy the Lipschitz and linear growth conditions \cite{gogoi2022computational}. For the solution of Eq.~\eqref{eq:genSDE}, numerical integration schemes of different orders of accuracy can be constructed by incorporating Ito's lemma in Eq.~\eqref{eq:genSDE} in conjunction with integral form of Taylor series expansion \cite{gogoi2022computational,kloedenplaten1992}.
Considering all the first and second order multiple stochastic integrals (MSIs) along with one of the MSIs of multiplicity 3 from the Ito-Taylor expansion, the $k^{th}$ component of Ito-Taylor 1.5 discretization for the solution to Eq.~\eqref{eq:genSDE} is written as \cite{gogoi2022computational},
\begin{align}
\label{eq:generalTaylor}
y^{n + 1}_k = \,\, & y^n_k + {a_k}\Delta t + \sum\limits_{j = 1}^m {{b_{k,j}}\Delta {B^j}}  + \frac{1}{2}{\Im ^0}{a_k}{\left( {\Delta t} \right)^2} \notag \\ &+ \sum\limits_{j = 1}^m {\left( {{\Im ^0}{b_{k,j}}{I_{\left( {0,j} \right)}} + {\Im ^j}{a_k}{I_{\left( {j,0} \right)}}} \right)} \notag \\
& + \sum\limits_{{j_1},{j_2} = 1}^m {{\Im ^{{j_1}}}{b_{k,{j_2}}}{I_{\left( {{j_1},{j_2}} \right)}}} \notag \\ 
&+ \sum\limits_{{j_1},{j_2},{j_3} = 1}^m {{\Im ^{{j_1}}}{\Im ^{{j_2}}}{b_{k,{j_3}}}{I_{\left( {{j_1},{j_2},{j_3}} \right)}}},
\end{align}
where the expressions of the operators $\Im^0$ and $\Im^j$ evaluated on the drift and diffusion coefficients are given as \cite{gogoi2022computational},
\begin{subequations}
\label{eq:operators}
\begin{align}
{\Im ^0}\left( {g\left( {y\left( t \right)} \right)} \right) &= \frac{{\partial \left( {g\left( {y\left( t \right)} \right)} \right)}}{{\partial t}} + \sum\limits_{k = 1}^d {{a_k}\frac{{\partial \left( {g\left( {y\left( t \right)} \right)} \right)}}{{\partial {y_k}}}} \notag \\ 
&+ \frac{1}{2}\sum\limits_{k,l = 1}^d {\sum\limits_{j = 1}^m {{b^{k,j}}{b^{l,j}}\frac{{{\partial ^2}\left( {g\left( {y\left( t \right)} \right)} \right)}}{{\partial {y_k}\partial {y_l}}}} }, \\
{\Im ^j}\left( {g\left( {y\left( t \right)} \right)} \right) &= \sum\limits_{k = 1}^d {{b^{k,j}}\frac{{\partial \left( {g\left( {y\left( t \right)} \right)} \right)}}{{\partial {y_k}}}}.
\end{align}
\end{subequations}
The integral forms of the MSIs $I_{\left(0,j\right)}, I_{\left(j,0\right)}, I_{\left(j_1,j_2\right)}$ and $ I_{\left(j_1,j_2,j_3\right)}$ in Eq.~\eqref{eq:generalTaylor} can be phrased as \cite{gogoi2022computational},
\begin{subequations}
\label{eq:MSI12}
\begin{align}
&{{I_{\left( {0,j} \right)}} = \int\limits_t^{t + \Delta t} {\int\limits_t^{{s_1}} {\text{d}{s_2}\text{d}{B^j}\left( {{s_1}} \right)} } },  \\
&{{I_{\left( {j,0} \right)}} = \int\limits_t^{t + \Delta t} {\int\limits_t^{{s_1}} {\text{d}{B^j}\left( {{s_2}} \right)\text{d}{s_1}} } }, \\
&{{I_{\left( {{j_1},{j_2}} \right)}} = \int\limits_t^{t + \Delta t} {\int\limits_t^{{s_1}} {\text{d}{B^{{j_1}}}\left( {{s_2}} \right)\text{d}{B^{{j_2}}}\left( {{s_1}} \right)} } }, \\
&{{I_{\left( {{j_1},{j_2},{j_3}} \right)}} = \int\limits_t^{t + \Delta t} {\int\limits_t^{{s_1}} {\int\limits_t^{{s_2}} {\text{d}{B^{{j_3}}}\left( {{s_3}} \right)\text{d}{B^{{j_2}}}\left( {{s_2}} \right)\text{d}{B^{{j_1}}}\left( {{s_1}} \right)} } } },
\end{align}
\end{subequations}
where $\Delta t$ is the uniform step size of integration. For numerical implementation, the MSIs $I_{\left( {j,0} \right)}$ and $I_{\left( {0,j} \right)}$ in Eq.~\eqref{eq:MSI12} can be expressed as \cite{gogoi2022computational,kloedenplaten1992},
\begin{equation}
\label{eq:msigen}
\begin{array}{l}
{I_{\left( {j,0} \right)}} = \Delta {Z^j}, \\ 
{I_{\left( {0,j} \right)}} = \Delta {B^j}\Delta t - {I_{\left( {j,0} \right)}},
\end{array}
\end{equation}
with $\Delta B^j$ and $\Delta Z^j$ being the Brownian increments at each step \cite{gogoi2022computational}.
Considering commutative noise of the second kind \cite{gogoi2022computational}, the Ito-Taylor 1.5 discretization for multi-dimensional independent components of Wiener process is obtained as,
\begin{align}
\label{eq:Taylorsum}
y^{n + 1}_k = & y^n_k + {a_k}\Delta t + \sum\limits_{j = 1}^m {{b_{k,j}}\Delta {B^j}}  + \frac{1}{2}{\Im ^0}{a_k}{\left( {\Delta t} \right)^2} \notag \\ 
&+ \sum\limits_{j = 1}^m {\left( {{\Im ^0}{b_{k,j}}\left( {\Delta {B^j}\Delta t - \Delta {Z^j}} \right) + {\Im ^j}{a_k}\Delta {Z^j}} \right)} \notag \\ 
&+ \frac{1}{2}\sum\limits_{j = 1}^m {{\Im ^j}{b_{k,j}}\left( {{{\left( {\Delta {B^{{j_1}}}} \right)}^2} - \Delta t} \right)} \notag \\
&+ \sum\limits_{{j_1} = 1}^m {\sum\limits_{{j_2} = 1}^{{j_1} - 1} {{\Im ^{{j_1}}}{b_{k,{j_2}}}\Delta {B^{{j_1}}}\Delta {B^{{j_2}}}} } \notag \\
&+ \frac{1}{2}\sum\limits_{\scriptstyle{j_1},{j_2} = 1\hfill\atop
\scriptstyle\;{j_1} \ne {j_2}\hfill}^m {{\Im ^{{j_1}}}{\Im ^{{j_2}}}{b_{k,{j_2}}}\Delta {B^{{j_1}}}\left( {{{\left( {\Delta {B^{{j_2}}}} \right)}^2} - \Delta t} \right)} \notag \\ 
&+ \sum\limits_{{j_1} = 1}^m {\sum\limits_{{j_2} = 1}^{{j_1} - 1} {\sum\limits_{{j_3} = 1}^{{j_2} - 1} {{\Im ^{{j_1}}}{\Im ^{{j_2}}}{b_{k,{j_3}}}\Delta {B^{{j_1}}}\Delta {B^{{j_2}}}\Delta {B^{{j_3}}}} } } \notag \\
&+ \frac{1}{2}\sum\limits_{j = 1}^m {{\Im ^j}{\Im ^j}{b_{k,j}}\left( {\frac{1}{3}{{\left( {\Delta {B^j}} \right)}^2} - \Delta t} \right)\Delta {B^j}}.
\end{align}
For the most general case, i.e., for multiplicative noise, the numerical discretization in Eq.~\eqref{eq:Taylorsum} has a strong order of accuracy of 1.5 \cite{kloedenplaten1992}. For the equations of motion in Eqs.\eqref{eq:motion21}--\eqref{eq:motion23}, consider the state-space representation as,
\begin{equation}
\label{eq:statespace}
\left. {\begin{array}{*{20}{c}}
	{{y_1} := \theta},\\
	{{y_2} := {\dot{\theta}}},
	\end{array}} \right|\left. {\begin{array}{*{20}{c}}
	{{y_3} :=  \phi},\\
	{{y_4} := {\dot{\phi}}},
	\end{array}} \right| \begin{array}{*{20}{c}}
	{{y_5} := \psi},\\
	{{y_6} := {\dot{\psi}}}.
	\end{array}
\end{equation}
Accordingly, we express Eqs.\eqref{eq:motion21}--\eqref{eq:motion23} in the form of an Ito SDE (Eq.~\eqref{eq:genSDE}) as follows,
\begin{align}
    \label{eq:Itoform}
    \text{d}y_1 &= y_2 \text{d}t, \notag \\ 
    \text{d}y_2 &= \left(-\frac{I_3}{I_1}y_4y_6 - \gamma_1y_2 - \Omega^2_0y_1\right)\text{d}t + \frac{c_1}{I_1} \text{d}\eta_{\theta}, \notag \\
    \text{d}y_3 &= y_4 \text{d}t, \notag \\
    \text{d}y_4 &= \left(\frac{I_3}{I_1}y_2y_6 - \gamma_1y_4 - \Omega^2_0y_3\right)\text{d}t + \frac{c_1}{I_1} \text{d}\eta_{\phi}, \notag \\
    \text{d}y_5 &= y_6 \text{d}t, \notag \\
    \text{d}y_6 &= \left(-\gamma_3y_6 - \frac{1}{2}\Omega^2_{\psi}\sin{2y_5}\right)\text{d}t + \frac{c_2}{I_3} \text{d}\eta_{\psi},
\end{align}
where $c_1 = \sqrt{2I_1\gamma_1k_BT}$ and $c_2 = \sqrt{2I_3\gamma_3k_BT}$. The drift and diffusion coefficients can now be identified from Eq.~\eqref{eq:Itoform} which are then used in Eq.~\eqref{eq:Taylorsum} together with Eq.~\eqref{eq:operators} to evaluate the trajectories of the state variables defined in Eq.~\eqref{eq:statespace}. Based on Eqs.~\eqref{eq:statespace} and \eqref{eq:noiseterms}, $k$ and $m$ in Eq.~\eqref{eq:Taylorsum} are $6$ and $3$ respectively.


\section{Results}

\begin{figure}
    \centering
    \includegraphics[width=0.9\linewidth]{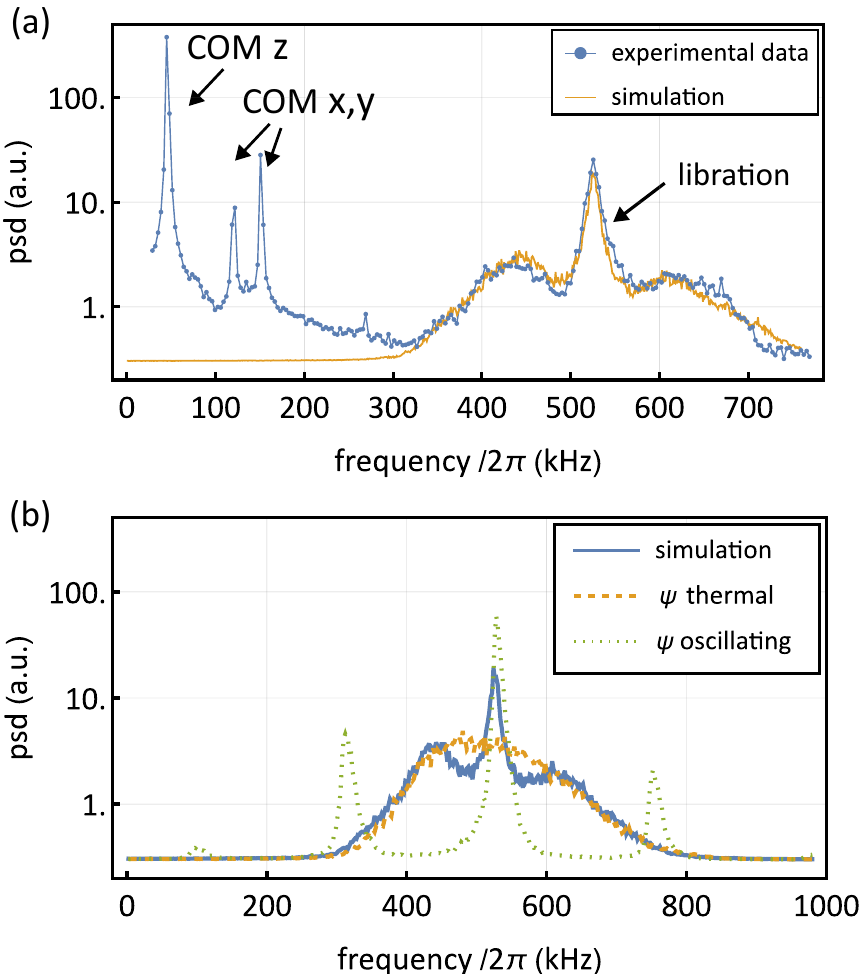}
    \caption{Power spectral densities. {\bf (a):} Experimental and simulated $S_{\phi\phi}(\omega)$. The libration motion appears as a peak with shoulders centered at \qty{525}{\kilo\hertz}. Cross-talk between detection channels results in COM oscillations visible in the experimental PSD at low frequencies. {\bf (b):} Simulated $S_{\phi\phi}(\omega)$ with a sinusoidal confinement potential for $\psi$, as per Eqs.\eqref{eq:motion21}--\eqref{eq:motion23}. The dashed line shows the simulated $S_{\phi\phi}(\omega)$ in the absence of confinement for $\psi$ ($\Omega_\psi = 0$), where the $\psi$ motion is purely thermal. The dotted line corresponds to a scenario where the confinement potential for $\psi$ is purely harmonic. The simulation time in all cases is \qty{100}{\milli\second}.}
    \label{fig:spectra}
\end{figure}

\subsection{Reference experimental data}

We employed the stochastic simulation described previously to explain the rotational dynamics of the nanodumbbell studied experimentally in Ref.~\cite{zielinska2024PRL}. The nanodumbbell in question consists of two silica spheres, each with a nominal diameter of \qty{143}{\nano\meter}. It is confined using linearly polarized optical tweezers operating at a wavelength of \qty{1550}{\nano\meter}, with an optical power of \qty{700}{\milli\watt}, and focused through a lens with a numerical aperture (NA) of 0.8.
The particle is levitated at room temperature inside a vacuum chamber maintained at a pressure of \qty{1.5}{\milli\bar}. The angular trajectory $\phi(t)$ is inferred from polarization fluctuations in the trapping beam, introduced by light scattered from the particle (using forward-scattering scheme)~\cite{vanderLaan2021PRL}. The resulting libration power spectral density, shown in Fig.~\ref{fig:spectra}(a), is related to $\phi(t)$ as follows~\cite{li2017probability}:

\begin{equation}
    S_{\phi\phi}(\omega)= \lim_{T\rightarrow\infty} \frac{\langle|\tilde \phi_T (\omega)|^2\rangle}{T}, 
\end{equation}
where $\tilde\phi_T(\omega)=\frac{1}{2\pi}\int_{-T/2}^{T/2}dt \;\phi(t)e^{i\omega t}$ is a truncated Fourier transform and $\langle \cdot \rangle$ denotes the expectation value.

The libration mode $\phi$ is visible in Fig.~\ref{fig:spectra}(a) at \qty{525}{\kilo\hertz}, with lineshape characterized by a sharp central peak flanked by shoulders spaced approximately \qty{100}{\kilo\hertz} from the main peak. Although the measurement scheme is sensitive only to the $\phi$ angle~\cite{Tebbenjohanns2022PRA}, the libration detector also registers COM peaks due to cross-talk between the libration and COM detection channels. The COM motion along the beam propagation axis ($z$) is observed at \qty{50}{\kilo\hertz}, while the transverse COM modes (corresponding to particle oscillations along $x$ and $y$) appear near \qty{150}{\kilo\hertz}.

\subsection{Simulated libration power spectral density}

The best-fit power spectral density $S_{\phi\phi}(\omega)$, obtained by simulating a \qty{100}{\milli\second} trajectory of all three angular degrees of freedom using our stochastic model and simulation framework, is shown alongside the experimental data in Fig.~\ref{fig:spectra}(a). The following parameters were fixed: the libration frequency $\Omega_0 = 2\pi \times \qty{525}{\kilo\hertz}$, the damping rates $\gamma_1 = \gamma_3 = 2\pi \times \qty{1.5}{\kilo\hertz}$, and the aspect ratio of the moments of inertia, $I_3 / I_1 = 0.66$, consistent with a nanodumbbell characterized by a length-to-diameter ratio of 1.8. The free parameters in the simulation are the particle size  characterized by the moment of inertia \( I_1 \), with \( I_3 \) determined by a fixed aspect ratio and the confinement strength of the \( \psi \) degree of freedom, specified by the frequency \( \Omega_\psi \). To align the simulation with the experimental spectrum, the simulated power spectral density was additionally scaled by a calibration factor and offset by a constant baseline.

The simulation parameters yielding the best agreement with experimental data were \( \Omega_\psi = 2\pi \times \qty{220}{\kilo\hertz} \) and \( I_1 = \qty{2e-33}{\kilogram \cdot \meter^2} \). The value of \( \Omega_\psi \) can be attributed to intensity gradient torques~\cite{kamba2023nanoscale} and to the longitudinal field component arising from strong focusing~\cite{novotny2012principles}. However, the inferred moment of inertia \( I_1 \) is roughly an order of magnitude smaller than expected based on the nominal particle size and the density of silica. The value of \( I_1 \)  determines the magnitude of thermal fluctuating torques in the simulation, according to  Eq.~\eqref{eq:Itoform}. Possible explanations for this discrepancy include lower effective density of the particle and/or the presence of additional noise sources in the experiment, such as laser relative intensity noise (RIN), elevated local temperature, or coupling between the rotational and COM motion of the particle. The fact that the particle behaves as a much smaller one can be already deduced from the experimental data in Fig.~\ref{fig:spectra}(a): the observed sidebands appear approximately \( \delta \Omega = 2\pi \times \qty{180}{\kilo\hertz} \) apart. To explain the splitting of libration into two modes with frequencies \( \delta \Omega \) apart, the particle would need to be spinning at 
\( \dot{\psi} = \delta \Omega \, I_1 / I_3 \approx 2\pi \times \qty{273}{\kilo\hertz} \)~\cite{seberson2019parametric}. 
In contrast, using the moment of inertia obtained from the nominal particle diameter and silica density, the expected spinning frequency (assuming the kinetic energy of \( \frac{1}{2}k_{\mathrm{B}}T \) associated with the long axis rotation) would be only \( \dot{\psi} \approx 2\pi \times \qty{87}{\kilo\hertz} \).

\begin{figure}
    \centering
    \includegraphics[width=0.9\linewidth]{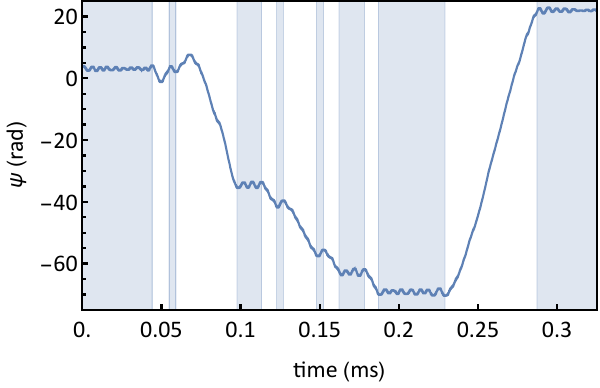}
    \caption{Fragment of a simulated trajectory \( \psi(t) \). Shaded regions indicate periods of libration, where the intermediate axis oscillates around the \( z \)-direction. Unshaded regions correspond to the rotational regime, during which the particle spins around its long axis.}
    \label{fig:trajectory}
\end{figure}
\begin{figure}
    \centering
    \includegraphics[width=0.9\linewidth]{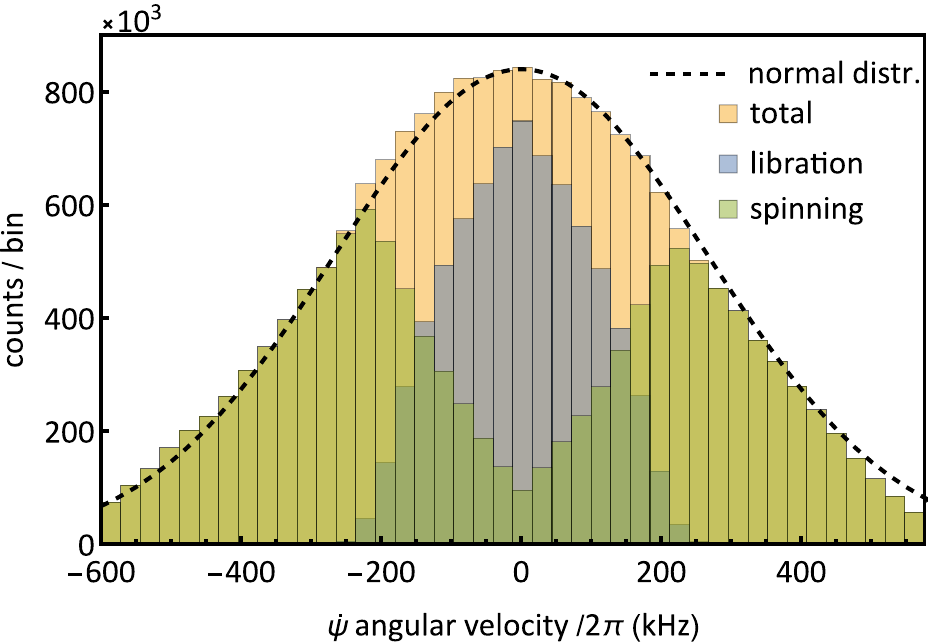}
    \caption{Histogram of the angular velocity distribution \( \dot{\psi} \). The yellow region represents the total distribution over the full \qty{100}{\milli\second} simulation. The green region shows the distribution during rotational motion, while the blue region corresponds to periods of libration. The dashed line indicates a fitted normal distribution.}
    \label{fig:histograms}
\end{figure}

\subsection{Librating and rotating regime of the long axis rotation}

For the simulation parameters that best reproduce the experimental power spectral density, the depth of the potential confining the angle \( \psi \) given by Eq.~\eqref{eq:psipotential} (yielding \( \frac{1}{4} I_3 \Omega_\psi^2 \)), is on the order of \( \frac{1}{2} k_B T \). Due to thermal fluctuations the particle alternates between two regimes: during certain intervals, the intermediate axis is trapped by the confining potential, resulting in oscillatory dynamics of \( \psi \); at other times, its energy exceeds the potential depth, allowing \( \psi \) to escape confinement and the particle to rotate freely about its long axis. This behaviour is clearly visible in Fig.~\ref{fig:trajectory}.

The effects of the two regimes of \( \psi \) motion are reflected in the shoulders flanking the libration peak. Figure~\ref{fig:spectra}(b) additionally shows what the power spectral density \( S_{\phi\phi}(\omega) \) would look like in each regime: if the angle \( \psi \) rotated freely, the peak broadens into a rounded, Gaussian-like shape; if \( \psi \) were confined in a harmonic potential with natural frequency \( \Omega_\psi \), the peak becomes a Lorentzian with symmetric sidebands. These sidebands arise because the term \( \dot{\theta} \dot{\psi} \) in Eq.~\eqref{eq:motion22} has an effect similar to a torque oscillating at frequencies \( \Omega_0 \pm \Omega_\psi \). Note that in the case of a sinusoidal (i.e., anharmonic) potential, the oscillations of the angle \( \psi \) span a range of frequencies extending up to a maximum of \( \Omega_\psi \), which results in broadened sidebands. 

Fig.~\ref{fig:histograms} further illustrates how the combination of rotational and librational regimes of the \( \psi \) angle gives rise to the shouldered peak structure observed in the libration PSD. When the particle rotates freely around its long axis, it cannot do so with low \( \dot{\psi} \) values- if \( \dot{\psi} \) becomes too small, the intermediate axis becomes confined by the potential, transitioning the particle back into the libration regime. As a result, thermal spinning (i.e., free rotation) occurs predominantly at high \( \dot{\psi} \) velocities, leading to shoulders structure. Despite the transition between regimes, the total velocity distribution remains approximately Gaussian, following Maxwell-Boltzmann distribution. Analysis of the simulated \qty{100}{\milli\second} trajectory shows that in our case, the $\psi$ degree of freedom spends approximately 53\% of the time in the librating regime and 47\% in the rotating regime.

To summarize, we identified that the phenomenon responsible for the observed libration peak with shoulders in optically trapped nanodumbbells is the long-axis rotation transitioning between librating and rotating regimes. Similar rotation-libration transitions have been studied for short-axis rotation~\cite{kuhn2017full,Bellando22PRL}.

\begin{figure}
    \centering
    \includegraphics[width=0.8\linewidth]{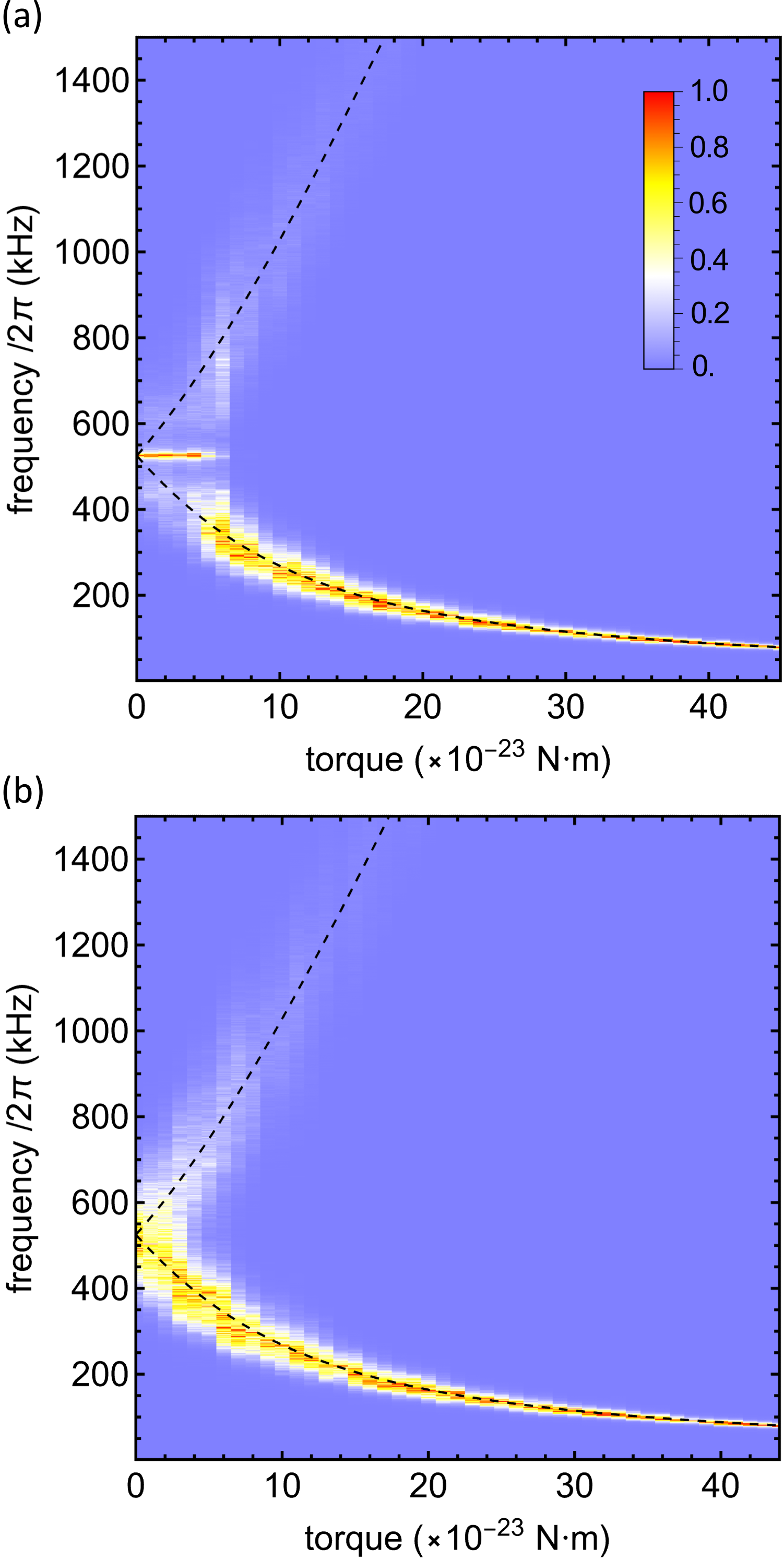}
    \caption{Libration PSD in arbitrary units, showing the effect of the spinning torque.  Black dashed lines indicate the predicted precession and nutation frequencies calculated from Eq.~\eqref{eq:precnut}. 
\textbf{(a)}: Precession (low-frequency, narrow) and nutation (high-frequency, broad) modes emerge once the applied spinning torque exceeds a threshold value.  
\textbf{(b)}: Same as (a), but without the confining potential ($\Omega_\psi=0$).
 }
    \label{fig:torque}
\end{figure}

\subsection{Effect of an additional torque spinning the particle along its long axis}

To gain further insight into the effect of \( \psi \) confinement, we simulated the experimental conditions described in Fig.~3(a) of Ref.~\cite{zielinska2024PRL}. In that experiment, an additional beam applied a spinning (non-conservative) torque $\tau$ along the optical tweezers’ polarization direction, causing the dumbbell to rotate around its long axis. The long-axis spinning speed increased until reaching a steady-state value of \( \dot{\psi}_0 = \tau / (I_3 \gamma_3) \). A simple theoretical model predicts that the originally degenerate libration modes hybridize into precession and nutation modes with frequencies given by
\begin{equation}
 \Omega_{1/2} = \sqrt{\Omega_0^2 + g^2} \pm g \;,
 \label{eq:precnut}
 \end{equation}
 where \( g = (I_3 / 2 I_1) \dot{\psi}_0 \)~\cite{zielinska2024PRL}. However, this model did not predict the linewidths of these modes and failed to accurately describe the particle’s behavior under small torque, where a threshold for mode splitting was observed. Notably, the libration mode retained its characteristic shoulder-like peak and only transitioned to distinct precession and nutation modes once the applied torque exceeded a certain minimum value.

Our stochastic modeling and simulation, in turn, shows that the intermediate axis sinusoidal potential described by Eq.~\eqref{eq:psipotential} is responsible for the threshold. The spinning torque has been modelled by including an additional torque term \( \tau \) in Eq.~\eqref{eq:motion23}, so that it becomes
\begin{equation}
I_3 \ddot{\psi}+ I_3 \gamma_3 \dot{\psi}+\frac{1}{2}I_3 \Omega_\psi^2 \sin{2\psi}=\tau+\tau_{{\rm fl},\psi},
\end{equation}
while Eqs.~\eqref{eq:motion21} and~\eqref{eq:motion22} remain unchanged.
 
The simulation results shown in Fig.~\ref{fig:torque}(a) qualitatively reproduce both the threshold behavior and the experimentally observed linewidths of the precession and nutation modes. The threshold corresponds to the minimum non-conservative torque required to overcome the confining potential of the intermediate axis, enabling the dumbbell to spin persistently in the direction of the applied torque. For comparison, Fig.~\ref{fig:torque}(b) presents simulation results without the \( \psi \) confinement potential, i.e., with \( \Omega_\psi = 0 \). In this case, no threshold is observed, and the precession and nutation frequencies are accurately described by Eq.~\eqref{eq:precnut}.

\section{Conclusion}

This paper presented a stochastic simulation framework and a related model to 
reconstruct the complex dynamics of long-axis rotation (\( \psi \))- a degree of freedom not directly accessible through experiments, by analyzing the readily detectable short-axis rotation (\( \phi \)). The proposed simplification of the full nonlinear rotational dynamics
was demonstrated to be sufficient 
for explaining and modelling benchmark experimental data Refs.~\cite{Zielinska2023PRL,zielinska2024PRL}. The proposed model and framework can be  readily extended to include the remaining nonlinear terms from the complete equations of motion, as well as couplings to the COM degrees of freedom~\cite{rademacher2025roto,kamba2023revealing}. Consequently, it has the potential to  be relevant in a generalised sense to understand the motion of high-aspect-ratio particles subject to strong coupling between rotational and translational motion, as well as for accurately reconstructing dynamics in engineered optical potential landscapes created by strongly focused structured light~\cite{PhysRevA.104.013516, hu2023structured}.
 
The proposed Ito-Taylor expansion based stochastic integration approach can generate a complete trajectory of all three rotational degrees of freedom- including key nonlinearities within seconds on a standard laptop. This computational efficiency makes the method a practical and powerful tool for modeling and validation of rotational optomechanics experiments, along with development of new applications around high-precision devices measuring kinematic outputs. 

\section{Acknowledgements}


Joanna Zielinska acknowledges the support by Challenge-Based-Research project with the ID \text{CI\_EIC\_CLS\_S\_116} at Tecnológico de Monterrey. JZ also acknowledges the Photonics Laboratory at ETH Zurich, where she was previously based and where the experiments reported here were carried out. Ankush Gogoi and Vikram Pakrashi acknowledge the support of Research Ireland NexSys (21/SPP/3756) and Harmoni 22FFP-P11457.


\bibliographystyle{unsrt}
\bibliography{sample}

\end{document}